\documentstyle[12pt,epsf]{article}

\begin{document}

\title{Accuracy of Trace Formulas}
\date{}
\author{Arul Lakshminarayan  \\
{\sl Physical Research Laboratory,}\\
{\sl Navarangapura, Ahmedabad,  380009, India.}}
\maketitle

\newcommand{\newc}{\newcommand}
\newc{\beq}{\begin{equation}}
\newc{\eeq}{\end{equation}}
\newc{\beqa}{\begin{eqnarray}}
\newc{\eeqa}{\end{eqnarray}}
\newc{\pa}{\partial}
\newc{\bom}{\boldmath}
\newc{\btd}{\bigtriangledown}
\newc{\rarrow}{\rightarrow}
\newc{\bea}{\begin{array}}
\newc{\eea}{\end{array}}
\newc{\mb}{\mbox}
\newc{\tm}{\times}
\newc{\kt}{\rangle}
\newc{\br}{\langle}
\newc{\ra}{\Rightarrow}
\newc{\longra}{\longrightarrow}
\newc{\longla}{\longleftarrow}
\newc{\ts}{\textstyle}
\newc{\noi}{\noindent}
\newc{\f}{\frac}
\newc{\eps}{\epsilon}
\begin{abstract}
	Using quantum maps we study the accuracy of semiclassical
trace formulas. The role of chaos in improving the semiclassical
accuracy, in some systems, is demonstrated quantitatively.  However,
our study of the standard map cautions that this may not be most
general.  While studying a sawtooth map we demonstrate the rather
remarkable fact that at the level of the time one trace even in the
presence of fixed points on singularities the trace formula may be
exact, and in any case has no logarithmic divergences observed for the
quantum bakers map.  As a byproduct we introduce fantastic periodic
curves akin to curlicues.

\vspace{.5cm}


\vspace{.5cm}
\noindent {\sl Submitted to} Pramana, Journal of Physics, {\sl (Special Issue
on Nonlinearity and Chaos in Physical Sciences.)} 

\end{abstract}
\newpage

\section{Introduction}

\hspace{.5in} 

	The trace formula of Gutzwiller [1] provides a window into the
semiclassical treatment of chaotic systems and is hence of
considerable importance in many areas of physics and chemistry. Yet
the understanding of this formula has been hampered by the complexity
of the classical task involved, and often the semiclassical effort is
greater than that for a full quantum solution which makes the
semiclassical methods either only of academic interest or as  tools
for explaining features of a full quantum solution. In this situation
simple models have played a key role just as has been the case with
the understanding of classical chaos, as evidenced for example by the 
study of the logistic map and the smale horseshoe. 
	
	Although the trace formula (and its relatives) are important,
the accuracy of these have not been rigorously established. The trace
formula being the first term in an asymptotic expansion, has many
approximations with apparently uncontrolled errors. The very many
applications of the stationary phase approximation, on functional and
on ordinary integrals obscures the nature of the errors. One ongoing
issue is the {\em time} at which we can expect the various
semiclassical approximations to go bad with a candidate the so called
log time which is of the order of $-\log(\hbar)/\lambda$, applicable
for chaotic systems with a Lyapunov exponent $\lambda$, being called
too pessimistic [2]. We note that although the
semiclassical formula for the time evolution (due to Van Vleck) has
been essentially known almost since the beginning of quantum mechanics
the nature of the errors is still a matter of interest, especially
since the trace formula depends on a Laplace transform of the Van
Vleck approximation [1].

	The quantum bakers map has been studied as a model of quantum
chaos as indeed it has many generic features of quantized chaotic
systems [3,4,5], however a rigorous
study of some of the simplest traces of the quantum problem [5] showed
logarithmic divergence from the expected Gutzwiller trace formula. The
genesis of this disturbing fact was attributed to a fixed point on the
line of discontinuity (corners) of the classical map. A baker without
corners was studied [6], and it was found that there were no
logarithmic divergences anymore. However we will in this note display
a map, the sawtooth map, which has a fixed point on the line of
discontinuity, but which has no logarithmic divergences. In fact we
will show that in certain circumstances the semiclassical trace can be
exactly the quantum trace.

	In pursuance of certain results in [5], we use the existing
models of quantum chaos to study the accuracy of the simplest trace of
the quantum problem. To this end we will in particular use the much
studied quantum standard map and the much less studied quantum
sawtooth map both in the settings of the phase space as a cylinder and
as a torus. These being maps derived from time periodic systems we
study simply the trace of the relevant unitary Floquet operators. In
section 2 the cylindrical phase space is used, while in section 3 the
toral phase space is used, we end with some comments in section 4. In
the Appendix we provide a self contained derivation of the ``trace
formula'' for a class of maps on the torus that includes the maps
studied below and all the so called kicked systems. While this is in
principle known, it may be useful to recollect it here, as the other
published proof of Tabor [17] is for the case of maps on the entire plane,
and is hence quite different. Those with little exposure to the trace
formula may peruse the Appendix first.

\section{The Cylindrical Phase Space}

	Consider the well known standard map $(q_{n+1}=q_{n}+T p_{n},
p_{n+1}=p_{n}+K \sin(q_{n+1}))$ defined on the cylinder $[0,2 \pi)
\times (-\infty,\infty)$ with periodic boundary conditions in $q$. The
first iterative equation is them to be evaluated with a modulo $2\pi$
rule. A simple change of variables from $p$ to $T p$ implies that the
relevant parameter is just one in number and is $k=KT$.  This area
preserving map and its quantization have been studied in extensive
detail as a model of chaos and its quantum manifestations and led to
such important discoveries as the quantum suppression of chaotic
diffusion. Some of these are reviewed in [7]. However its use
in understanding the Gutzwiller trace formula has been limited. The
work of Tabor [17] where he derived the relevant trace formula for quantum
maps on the plane did use the standard map but no rigorous or
extensive study has yet been attempted. The quantization of the
standard map is described by an infinite dimensional unitary matrix
with matrix elements in the momentum representation being 

\beq U_{n \,
m}\,=\, (-i)^{n-m} J_{n-m}(k/T \hbar) \, \exp(-i m^{2}\hbar T/2).
\eeq
	Here $J_{\nu}(z)$ is a Bessel function. As has been noted
earlier there are two relevant parameters quantum mechanically and we 
can take $T=1$ while we do not have the freedom to choose $\hbar=1$ anymore.
The trace of the quantum map is then given by
\beq
\mbox{Tr}( U) \, =\, J_{0}(k/\hbar) \sum_{n=-\infty}^{\infty} e^
{-i \hbar n^{2}/2} \,=\, J_{0}(k/\hbar) \Theta_{3}(0|-\hbar/2 \pi)
\eeq
where $\Theta_{3}$ is a Jacobi theta function. An application of the Poisson
formula enables us to rewrite the above as 
\beq
\mbox{Tr}( U) \, =\, \sqrt{\frac{2 \pi}{i \hbar}}
J_{0}(k/\hbar) \sum_{n=-\infty}^{\infty} \exp(2 \pi^{2} i  n^{2}/\hbar)
\eeq
Strictly speaking this sum need not exist and may without further
regularization be infinite. This is not due to  any peculiarities of
maps, and in fact this is the rule in quantum mechanics, the 
trace of the propagator could be infinite in real time. However this
will not worry us further as the semiclassical trace will also 
diverge exactly, allowing us to make meaningful relative error estimates.

Now we turn to a semiclassical evaluation of the quantum trace. We wish
to then write the sum in terms of the period one or fixed points of
the classical map.  The Lagrangian generating the standard map on the 
cylinder is 
\beq
L(q_{n},q_{n+1},l_{n})=(q_{n+1}-q_{n}+2 \pi l_{n})^{2}/2-k\cos(q_{n+1})
\eeq 
as $\pa L/\pa q_{n} \,=\, -p_{n}$ and $\pa L/\pa q_{n+1} \,=\,
p_{n+1}$.  $l_{n}$ is the winding number in the $q$ direction or $2
\pi l_{n}$ is subtracted by the modulo operation performed at time $n$
to keep the map on the cylinder.  The fixed points are determined by
the condition $(q_{n+1}=q_{n}, p_{n+1} =p_{n})$, and are therefore
simply given by $(q=0, p=2 \pi l)$ and $(q=\pi, p=2 \pi l)$, where $l$ runs
over the integers. We note that the number of fixed points are
infinite and that their positions do not depend on the classical chaos
parameter $k$. The denominator in the Gutzwiller trace formula (or its
versions for the case of quantum maps, they are essentially same) has
the factor $ \sqrt{\mbox{Tr}(J)-2}$, where $J$ is the classical
stability matrix associated with the periodic orbit. In the case of
standard map fixed points this is either $\sqrt{k}$ or $\sqrt{-k}$
depending on whether $q=0$ or $q=\pi$. 

The semiclassical approximation is of the form 
\beq
\mbox{Tr}(U)_{sc}=\sum_{\mbox{fixed points}} \exp(i
S/\hbar)/\sqrt{(Tr(J)-2)} 
\eeq 
where $S$ is the classical action and
since we are dealing with discrete time systems with the time interval
being taken as unity this is identical to the Lagrangian written
above. We get for the standard map on the cylinder the following semiclassical
approximation for the trace (see Appendix):
\beq
\mbox{Tr}(U)_{sc}=\sqrt{\frac{4}{i k}} \cos(\f{k}{\hbar} -\f{\pi}{4}) 
\sum_{l=-\infty}^{\infty} \exp(2 \pi^{2} i  l^{2}/\hbar)
\eeq
This then has to be compared with the Eq. (3). The asymptotic form
of $J_{0}(x)$ for large arguments is [8]
\[
J_{0}(x) \sim \sqrt{\frac{2}{\pi x}} [ \cos(x-\frac{\pi}{4}) \, P(x) \, - \, 
\sin(x-\frac{\pi}{4})\, 
Q(x)]
\]
where 
\[ 
P(x)=1-9/(128 x^{2})+\dots \,\,\mbox{and} \, \, Q(x)=-1/(8 x)+\dots
\]
where the neglected terms are at most of order $x^{-3}$. 
We then derive that
\beq
\frac{\mbox{Tr}(U)}{\mbox{Tr}(U)_{sc}} = P(\frac{k}{\hbar})-
\tan(\frac{k}{\hbar}-\frac{
\pi}{4}) \, Q(\frac{k}{\hbar}).
\eeq

Even this simple result is instructive. It indicates spectacular
failures of the trace formula when the ratio $k/\hbar$ passes through
$\pi/4+(2 m+1) \pi/2$, where $m$ is a positive integer.  At these
points the semiclassical trace becomes zero and the exact
quantum trace is also small indicating that such a failure will not be
serious for the evaluation of the spectra. But it does suggest that
the ratio of the quantum to the semiclassical need not be bounded in
the entire range of the parameters. This divergence of the ratio is
thus not due to any coalescing stationary points or bifurcations in
general; as we have already noted the fixed points are independent of
the chaos parameter $k$. Second, it is clear that for a larger $k$ or
chaos the semiclassical trace is going to be a better approximation.
The authors in [2] have observed this, and we see that the simplest
quantum trace provides a quantitative evidence of this. To compare the
absolute error rather than the relative, we have to deal with the
Jacobi theta function which occurs identically both in the quantum and
in the semiclassical traces, and as this can depend critically on the
rationality of $\pi/\hbar$ we do not pursue this here, except note
that in unbounded phase spaces such as the plane and the cylinder the
absolute value of the traces might actually diverge. 

The case of the standard map has been extensively studied but not so
the quantum mechanics of the sawtooth map, which is in many ways
simpler because it is a piecewise linear map for which it is easy to
extract classical information. Consider the map $(q_{n+1}=q_{n}+T
p_{n}, p_{n+1}=p_{n}+K (q_{n+1}-\pi))$ defined on the cylinder $[0,2
\pi) \times ( -\infty,\infty)$, which is completely chaotic for
positive $K$. This can be considered as a simple harmonically kicked
particle in one dimension. We may view the system as an otherwise free
particle on a unit circle being subjected to a kick whose strength is
proportional to the physical angle and which has a single
discontinuity when the angle is zero. The trace of the quantum
propagator may be written down just as in the case of the standard map
($T=1)$with the result that 
\beq \mbox{Tr}(U)\,=\, \frac{1}{\sqrt{K}}
\mbox{Erf} \left( e^{-i \pi/4} \sqrt{ \f{K}{2 \hbar}} \right) \, 
\sum_{n=-\infty}^{\infty}
\exp(2 \pi^{2} i n^{2}/\hbar),  
\eeq 
where Erf is the error function.

The semiclassical trace is also found quite easily, because the fixed points 
are at $(q=\pi,p=2 \pi l)$, $l$ being any integer. The classical action is 
similar to that of the standard map with the potential term being replaced by
a periodicised quadratic function. Thus we get 	
\beq 
\mbox{Tr}(U)_{sc}=\f{1}{\sqrt{K}} \,\sum_{n=-\infty}^{\infty}
\exp (2 \pi^{2} i n^{2}/\hbar),
\eeq
and the ratio of the quantum to semiclassical simply becomes
\beq
\f{\mbox{Tr}(U)}{\mbox{Tr}(U)_{sc}}\,=\, \mbox{Erf} \left(e^{-i \pi/4} 
\sqrt{\f{K}{2 \hbar}}\right).
\eeq
Inserting the asymptotic form for the error function we get the
relative error to be 
\beq
\left | 1-\f{\mbox{Tr}(U)}{\mbox{Tr}(U)_{sc}} \right | \approx 
\sqrt{\f{2 \hbar}{\pi K}}.
\eeq
The neglected terms are at most of order $(\hbar/K)^{3/2}$.

In this map also we see the seemingly general result that the larger
the chaos (larger $K$) the semiclassical accuracy is better. The
precise way in which this occurs for this map is like $K^{-1/2}$. The
difference between this map and the standard map is that the semiclassical
approximation seems more well behaved and the sudden decline in the
accuracy in the case of the standard map, due to the mismatch between
the zeros of the quantum and the semiclassical is absent here.
However in the regions where the semiclassical (and quantum) traces
are not very close to zero, the standard map would perform better as
the errors are of order $\hbar/k$ while for the sawtooth map the
errors are of order $(\hbar/k)^{1/2}$.  We will see that surprisingly
many of our observations will remain valid in the toral setting as
well.

\section{The Toral Phase Space}

	Two dimensional maps on the torus (periodic boundary
conditions in {\em both} position and momentum) have been studied
extensively, from the standard map to the cat map of Arnold. Their
quantizations have also been investigated by many [3,7,9],
as this is a legitimate model for a bound Hamiltonian system with two
degrees of freedom, the minimum needed for chaos.  Unlike the case of
the cylindrical phase space where scattering may take place, sheer
topology now prohibits this on the torus. The act of taking periodic
boundary conditions on the momenta is reminiscent of a similar
procedure in condensed matter physics which has also provided an
interesting two dimensional map, namely the kicked Harper
model. Quantum mechanics now is on a finite dimensional Hilbert space
and the we take the torus to be a unit torus (the dissected square is
of unit sides).  An alternate view of toral quantum maps is as the
case of ``resonance'' of the cylindrical map in which case the quantum
mechanics can be exactly reduced to that of a torus.  Position and
momentum take discrete values which we take to be $n/N$, where $N$ is
the dimension of the Hilbert space and $n\,=\,0 \ldots, N-1$.  Here
$N$ is required to be an {\em even integer} and is the inverse of the
Planck constant. Therefore the classical limit is $N \rarrow \infty$.

In this section we will first deal with the sawtooth map on the torus.
We will write the map a little differently for convenience as
$(q_{n+1}=(q_{n}+p_{n} )\mbox{mod} \, 1,\;\; q_{n+1}=(p_{n}+K q_{n+1}
) \mbox{mod} \, 1.)$.  The quantum map or propagator is given by 
\beq
U_{nm}\,=\,\frac{e^{-i \pi /4}}{\sqrt{N}} \exp(i \pi K m^{2}/N)\,
\exp(i \pi (n\,-\, m)^{2}/N).  
\eeq 
This then is the quantum
equivalent of the classical map restricted on a torus. It is a finite
unitary matrix as all quantum maps on compact phase spaces
are. Clearly it is one of the simplest quantum objects, having
elementary functions as matrix elements. Yet we find that it is rich
enough to reflect many essential features of the effect of classical
chaos on quantum systems. For more properties of a very similar
quantum map please refer [10].

The classical Lagrangian (or action) for this map is 
\beq
L(q_{n},q_{n+1},l_{n},m_{n})=(q_{n+1}-q_{n}+l_{n})^{2}/2 + K(q_{n+1}
(\mbox{mod}1))^{2}/2 -m_{n} q_{n+1}, 
\eeq 
where the integers $l_{n}$
and $m_{n}$ are winding numbers in the $q$ and $p$ directions
respectively. The fixed points are given by $(q=m/K, p=0)$ where
$m=0,1,2,\ldots,[K].$ The fixed point at the origin is in fact on the
line of discontinuity, and is quite like the bakers map and its
quantization. In that case it was noted that such singular periodic
orbits, ``half hyperbolic '', could give rise to unusual semiclassical
behaviour and that the quantum and semiclassical could diverge
semiclassically as $\log(\hbar^{-1})$.  A bakers map was constructed
to explicitly avoid this situation and a recovery of standard
expectations was realized [6].  However if we take the above
quantization as valid (since it follows canonical quantization closely
[11]) it will turn out that the fixed point on the discontinuity is
benign in the case of the sawtooth map, in that it does not lead to
semiclassically divergent terms. This may suggest that the unusual
nature of the bakers map originates either from the unusualness of the
very quantization procedure adopted or is peculiar to the bakers map.

Let us look at this case more closely, and take $0<K<1$, so that there
is a single fixed point at the origin which is also a point of
discontinuity.  A direct application of the Gutzwiller formula will be
incorrect in this case, but the rectification is obvious. The
stationary phase integral must not extend over the entire real line,
but must be restricted to $[0, \infty)$. Thus the semiclassical
approximation is then the approximation (quantum $\sim$ semiclassical)
\beq \f{e^{-i
\pi/4}}{\sqrt{N}} \, \sum_{n=0}^{N-1} \, \exp(i \pi K n^{2}/N) \, \sim
\, \f{1}{2 \sqrt{K}}.  
\eeq 
An estimate due to Wilton which we give below
bounds the difference as $N \rarrow \infty$, but this is not a good
estimate of the errors (for $0<K<1$), as these are of the order of the
sums themselves. An application of the Euler-Maclaurin summation rule
will give us the above approximation quite simply. But more important
is the observation that it is legitimate to replace summations with
integrals even in the presence of fixed points sitting on
discontinuities, a rule that was flagrantly violated in the case of
the bakers map. 

Let the errors be $\eps(K,N)\,=\,|\mbox{Tr} (U)-\mbox{Tr} (U)_{sc}|$,
that is the absolute value of the difference between the right hand
and the left hand side of the Eq. (14). We now observe some
remarkable number theoretic properties of these errors.  First we note
the surprising fact that if $N$ is a multiple of 4,
$\eps(1/2,N)=0$. In this case not only does the semiclassical and
quantum converge in the classical limit, they are exactly the same (at
the level of the trace), despite the presence of the fixed point on
the discontinuity!  We have noted before that the main feature
which makes the sawtooth map generic, in the sense that the
semiclassical is only an approximation, and in the sense of
eigenfunctions scarring and spectral random matrix properties is that the
incomplete Gauss sums that arise while compounding propagators do not
reduce to simple forms [10], unlike in the case of the cat maps ($K$ an
integer) where we get complete Gauss sums and has been investigated by
Hannay and Berry [9] and the extensive number theoretic properties were
explored by Keating [12]. Thus it is a bit interesting that an incomplete
Gauss sum can be done exactly and is the semiclassical approximation.

Consider the sum 
\beq \sigma_{N}=\sum_{n=0}^{N-1} \exp(i \pi
n^{2}/2N), 
\eeq 
which is incomplete because the period of $
s_{n}=\exp(i \pi n^{2}/2N) $ is $2N$ and the sum is only over the
first $N$ terms. First we note that $ s_{N/2 +m}=s_{N/2-m} \, \exp(i
\pi m), $ thus terms with odd $n$ do not matter as they cancel out of
the sum exactly.  Let \[ \Sigma_{2 N}\,=\, \sum_{n=0}^{2N-1} s_{n} \]
which is a complete Gauss sum [9]. But also note that if $N$ is a multiple
of 4, \[ s_{n+N}=(-1)^{n}s_{n}.  \] This then implies that \[
\sigma_{N} \,=\, \f{1}{2} \Sigma_{2N}\,=\, \f{1}{2}\sqrt{2N} e^{i
\pi/4}, \] therefore 
\beq \mbox{Tr} (U) \,=\, e^{-i \pi/4}
\sigma_{N}/\sqrt{N} = 1/\sqrt{2}.  
\eeq 
If $N$ is not divisible by 4, then $\eps(1/2,N)$ is no longer zero and
the semiclassical result is an approximation. Thus there are two
distinct classes when $K=1/2$ as far as the behaviour of the error is
concerned, This can be generalized and we observe that if $K=p/q$, is
a rational number, with relatively prime integers $p$ and $q$, then
there are as many classes as there are possible remainders of even
$N$, on division by $q$.  Within each class the error monotonically
goes to zero in the classical limit, $N \rarrow \infty$. We have
observed numerically that for large $N$ to an excellent approximation
\beq 
\eps(K,N)=C(K) N^{-1/2}.  
\eeq

Fig. 1(a) shows the case $K=2/5$, when there are five classes. The
constant $C(K)$ depends on both $K$ and the class, but we have
observed that $C(K)=C(1-K)$.  If $K$ is irrational, the number of
classes is infinite, there is no monotonic decrease of error, but the
general trend is to decrease, as is evident from Fig. 1(b), where we
have taken $K$ to be the most irrational number $\gamma=(\sqrt{5}-1)/2$, the
golden mean.  The number theoretical properties of the error is
of course a reflection of the sum itself, and a fuller study is thus
desirable. 

\begin{figure}[tbh]

\vskip 2mm
\hspace*{10mm}
\epsfxsize 8cm
\epsfysize 8cm
\epsfbox[18 150 590 716]{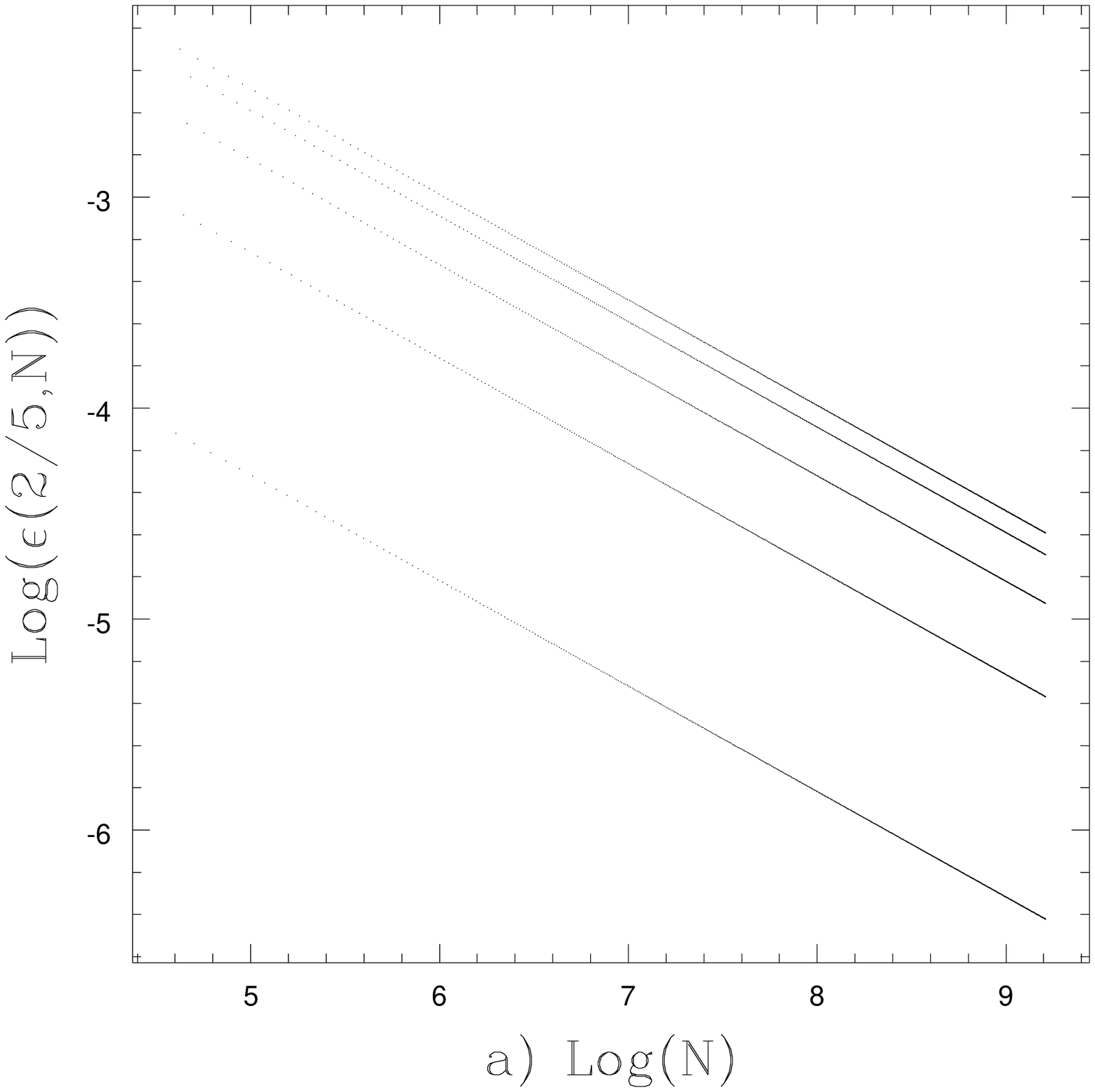}

\vspace*{-8cm}
\hspace*{9cm}
\epsfxsize 8cm
\epsfysize 8cm
\epsfbox[18 150 590 716]{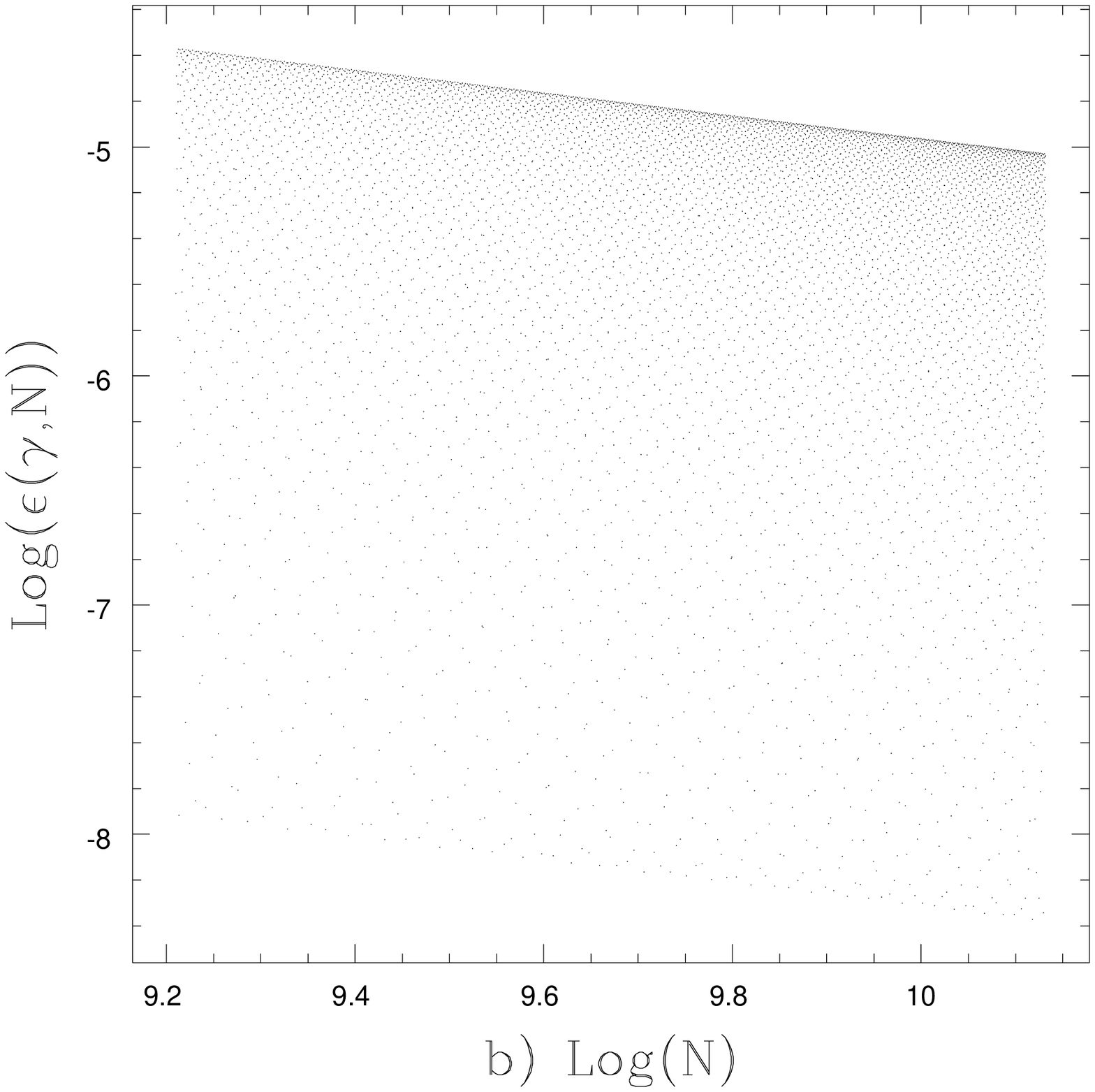}

\caption {The error as a function of inverse Planck's constant
$N$, for the sawtooth map with~ (a, left) $ K=2/5$;
(b, right) $K = \gamma $.}
\end{figure}
\vskip 3mm

We however shall now turn to the case when $K$ is larger
than one, and there are a number of fixed points. If $K$ is much larger than 
unity we need not correct for the corner fixed point as the errors incurred
will be of the order $K^{-1/2}$, which is anyway the order of the total error 
according to Hardy, Littlewood and Wilton, as quoted in [13]. 
The semiclassical approximation is now described by
\beq
\f{e^{-i \pi/4}}{\sqrt{N}}\sum_{n=0}^{N-1} \exp(i \pi K n^{2}/N) 
\, \sim \, \f{1}{\sqrt{K}} \sum_{n=0}^{[K]}\exp(-i \pi N n^{2}/K).
\eeq
Here $[K]$ is the integer value of $K$. Note that this approximation
clearly demonstrates the conjugacy between the Planck constant $(1/N)$
and the chaos parameter $K$. It is a good approximation if $K$ is
fixed and $N$ tends to infinity, namely the classical limit, but also
in the opposite limit of extreme ``chaos'', and small quantum
system. For example when $K$ tends to infinity for $N=2$, a two level
system. Undoubtedly, this approach to two level systems is to say the
least inadvisable, as the quantum problem is much easier to
solve. Nevertheless it throws up some interesting questions, because
what would one mean by extreme chaos in a two level system? Quantum chaos
for legitimacy in using the epithet chaos has always relied on the classical
limit, and in the extreme quantum limit the concept becomes nebulous.

The above approximation is derived from the Gutzwiller formula, using
the $[K]+1$ fixed point actions and the stability matrix. The fixed
point at the origin is included wrongly in the formula and we have
discussed it separately in the case $0<K<1$ above. Now we use an
estimate of Wilton, quoted in [13], 
to bound the error in this
approximation. This estimate is the inequality 
\beq
|\mbox{Tr}(U)-\mbox{Tr}(U)_{sc}| \, < \,
\f{a}{\sqrt{N}}+\f{b}{\sqrt{K}}, 
\eeq 
where $a$ and $b$ are constants
of order one whose values are slightly different from the quoted ones,
as the sawtooth sum is from $0$ to $N-1$, rather than 1 to $N$, but
this is a trivial complication. The essential conclusion is that the
error is bounded in $N$, and corroborates our conclusion that there is
no $\log(N)$ problems in the sawtooth map despite the presence of the
fixed point on the singularity. Also we once again notice that a
larger $K$ (larger chaos) implies a better semiclassical accuracy. In
Fig. 2 we show the results of a numerical computation of the error,
which shows that the error goes roughly as $K^{-1/2}$, with the slope
depending weakly on $N$ and $K$ and rather strongly on the fractional
part of $K$. Notice that at integer $K$ (cat maps) the approximation
is exact. We thus note that like its relative on the cylinder the
torus map error is also of the order of $K^{-1/2}$.

\begin{figure}[tbh]

\vskip 2mm
\hspace*{5cm}
\epsfxsize 8cm
\epsfysize 8cm
\epsfbox[18 150 590 716]{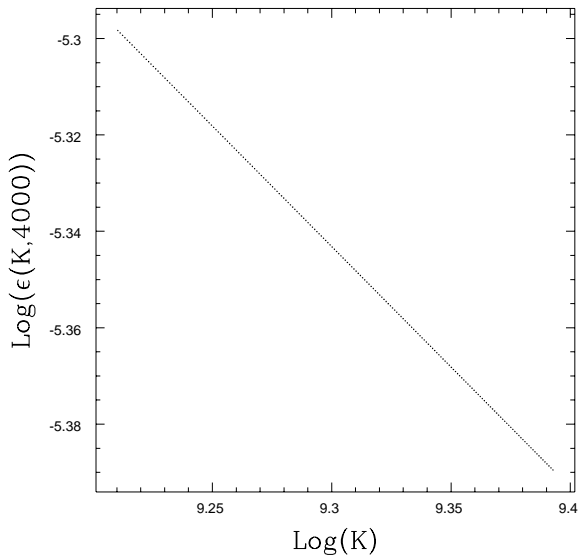}

\caption {The error as a function of the chaos parameter $K$,
for the sawtooth map with $N=4000$.}
\end{figure}
\vskip 3mm

We had earlier studied the sawtooth map on the torus with slightly
different boundary conditions [10], and came to essentially similar
conclusions, however in that case there was no fixed point on the
discontinuity. When the fixed points moved towards the singularity, at
the exact point of contact, the map turned into the continuous cat map
when the trace formula becomes exact.  We now turn to a smooth map on
the two torus, the standard map. Much studied in other contexts, it is
largely unstudied from the view of trace formulas, although it should
form a natural object of study, as it represents generic mixed
systems, and one recognizes the difficulties of such systems apropos
trace formulas.

We will consider the map once again on the unit torus. The map is now
$(q_{n+1}=(q_{n}+p_{n}) \mbox{mod} 1, \, p_{n+1}=(p_{n}+k \sin(2 \pi q_{n+1}))
\mbox{mod} 1)$. The Lagrangian (action) is 
\beq
L(q_{n},q_{n+1},l_{n},m_{n})=(q_{n+1}-q_{n}+l_{n})^{2}/2 -k\cos(2 \pi q_{n+1})
/2 \pi -m_{n} q_{n+1}.
\eeq 
The fixed points are easy to find and are at $(q_{m}=\f{1}{2
\pi}\arcsin(m/k),p_{m}=0)$, where m is an integer such that $|m|<
k$. Once again the location and number of the fixed points are very
much dependent on the the chaos parameter $k$ unlike the standard map
on the cylinder. Note that there are two fixed points associated with each $m$.

The quantum propagator is again given by ($N$ even throughout)
\beq
U_{n m}\, =\, \f{e^{-i \pi/4}}{\sqrt{N}} \exp(-i k N \cos(2 \pi n/N))
\exp(i \pi (n-m)^{2}/N).
\eeq
and the trace is the finite sum
\beq
\mbox{Tr}(U)\,=\, 
 \f{e^{-i \pi/4}}{\sqrt{N}} \sum_{n=0}^{N-1} \exp(-i k N \cos(2 \pi n/N)).
\eeq
The semiclassical approximation must approximate this sum using the
classical fixed points, their actions and their stabilities.  First
note that the quantum trace apart from the factor $e^{-i \pi/4}$ is
real. The sawtooth map traces were Gaussian sums, and partial sums of
these give rise to the ``curlicues'', that have been studied in detail
earlier and find applications from optics to quantum mechanics [13]. The
partial sums associated with the quantum trace of the standard map
ought to be interesting mathematical objects in their own right.

Here we present a preview of the intriguing class of periodic ``curves'' that 
this gives rise to. Consider the  sum 
\beq
\sigma_{N}(k,l)\,=\, \sum_{n=0}^{l}\exp(-i k N \cos(2 \pi n/N)).
\eeq
The quantum standard map trace is essentially $\sigma_{N}(k,N)$.  We
plot in Fig. 3 the real, $x(l)$, and imaginary, $y(l)$,
 parts of $\sigma_{N}(k,l)$ as an
argand diagram.  Strictly there are only as many points as terms in the 
partial sum, and as such do not define a curve, but the connections
showing the motion of the points in ``time'' $l$ is suggestive. 
Only the sum up to $l=N/2$ is relevant as beyond this
a reflection and periodic translation is the whole curve. Hence unlike
curlicues, these are periodic curves. A plethora of cornu spirals
interweave to conjure the extravagantly ornate features. Several ``eyes''
of cornu like spirals are connected in a remarkable manner. An
essential feature of these curves is their extreme susceptibility 
to small changes in $k$, especially at large $N$ values.
We cannot rule out the possibility of number theoretic features here
as well, and indeed we observe marked difference between the cases
when $k$ is an integer and when it is not. 

The periodic nature of the curves are easily understood, as $y(l+N/2)=
-y(l)$ and $x(l+N/2)=x(l)+A$, where $A\,=\,\sum_{n=0,N/2-1}\cos(-i k N
\cos(2 \pi n/N))$. Hence the period of the curves are $2A$.  As $N$ is
increased for fixed $k$ the curves tend to smoothen out more but with
the cornu spiral structures intact.
Also note that the quadratic approximation of the cosine
term in the exponential are the culicues studied earlier [13].

\vskip 2mm

\begin{figure}[tbh]
\hspace*{1cm}
\epsfxsize 15cm
\epsfysize 15cm
\epsfbox[18 150 590 716]{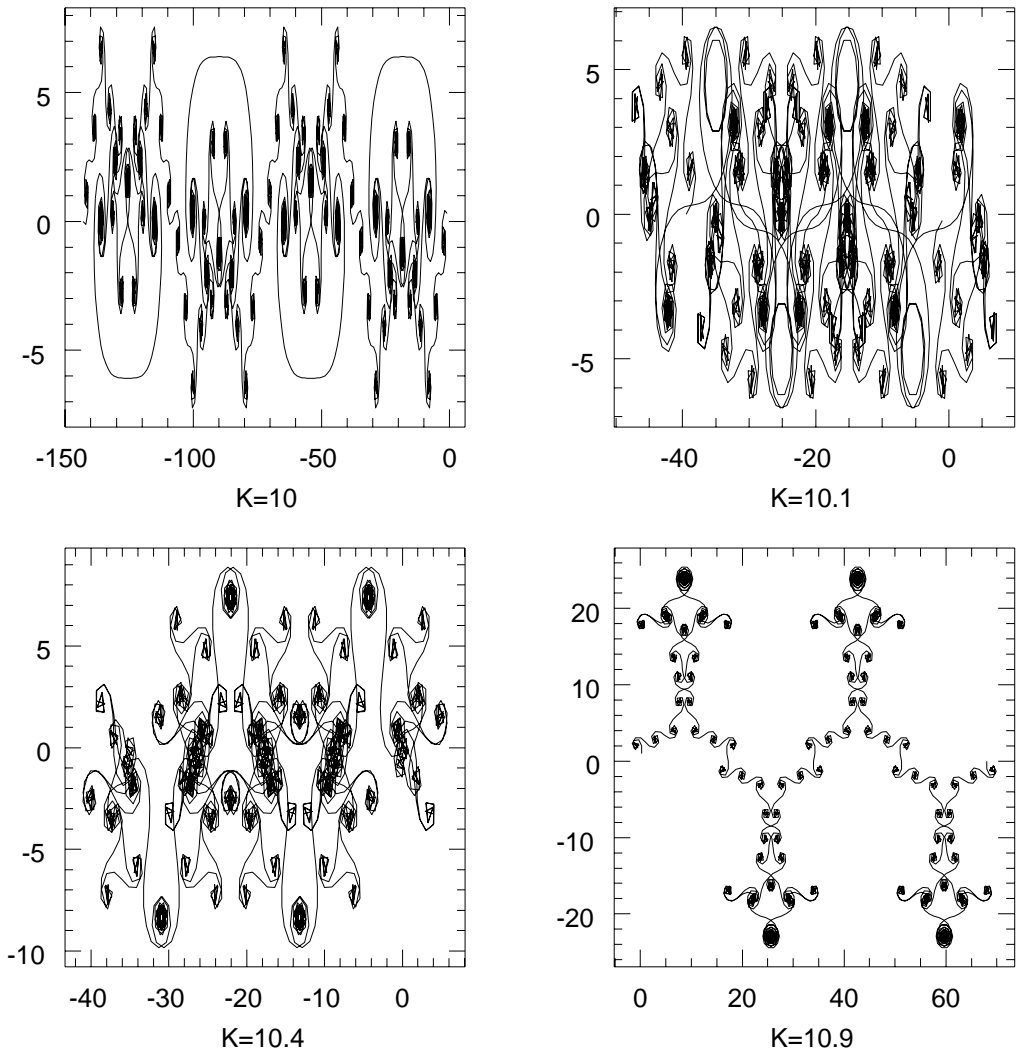}

\caption {A panel of the partial sums $\sigma_{1000}(K,l)$
(Eq. 23) for various $K$.}
\end{figure}
\vskip 3mm

\newpage

A detailed study of these
curves is however not pursued in this paper, and we turn once more to
the central theme, namely the accuracy of semiclassical
approximations.
Define
\beq
\theta_{N}(l)\,=\, N(\sqrt{k^{2}-l^{2}}\,+\, l \arcsin(l/k)).
\eeq
Then a careful application of the trace formula yields the 
semiclassical approximation
\beq
\mbox{Tr}(U)_{sc} \, =\, \f{e^{-i \pi/4}}{\sqrt{2 \pi}}
 \left [   \f{2 \sin (N k + \pi/4)}{\sqrt{k}} +
 \sum_{l=1}^{[k]} \f{4 \sin (\theta_{N}(l) + \pi/4)}
 {(k^{2}-l^{2})^{1/4}} \right],
\eeq
which is  real except for the factor $e^{-i \pi/4}$, just as the exact 
quantum trace in Eq. (22). Here the fixed point at the origin has been
treated separately as corresponding to each $|m|$ there are four fixed 
points except when $m=0$, which gives rise to two fixed points. The range 
of the inverse sine function is $[0, \pi/2]$. Note that this approximation 
can be derived independent of the trace formula, by first converting the  
 sum into an infinite sum of integrals via the Poisson summation formula, 
and of these integrals retaining only those which contribute to the 
stationary phase, and evaluating these integrals in the quadratic 
approximation. Indeed the trace formula in the context of maps 
on the torus is derivable by this procedure and is hence considerably 
simpler than the derivation for general Hamiltonian flows (see Appendix). 

First we note a drawback of the semiclassical approximation. Near
parameter values where bifurcations are going to create new periodic
points the semiclassical approximation becomes poor. Uniform
approximations may be made to smooth over these parameter values. For
instance in the case of the standard map, whenever $k$ passes an
integer value it creates four new fixed points via tangent
bifurcations, and the approximation to the trace becomes poor. This is
illustrated in Fig. 4(a).  However notice that the approximation is indeed
quite good even just away from these values. We will below retain the
approximation made above and keep away from points of bifurcations to
study various aspects of the error.

\begin{figure}[tbh]

\vskip 2mm
\hspace*{10mm}
\epsfxsize 8cm
\epsfysize 8cm
\epsfbox[18 150 590 716]{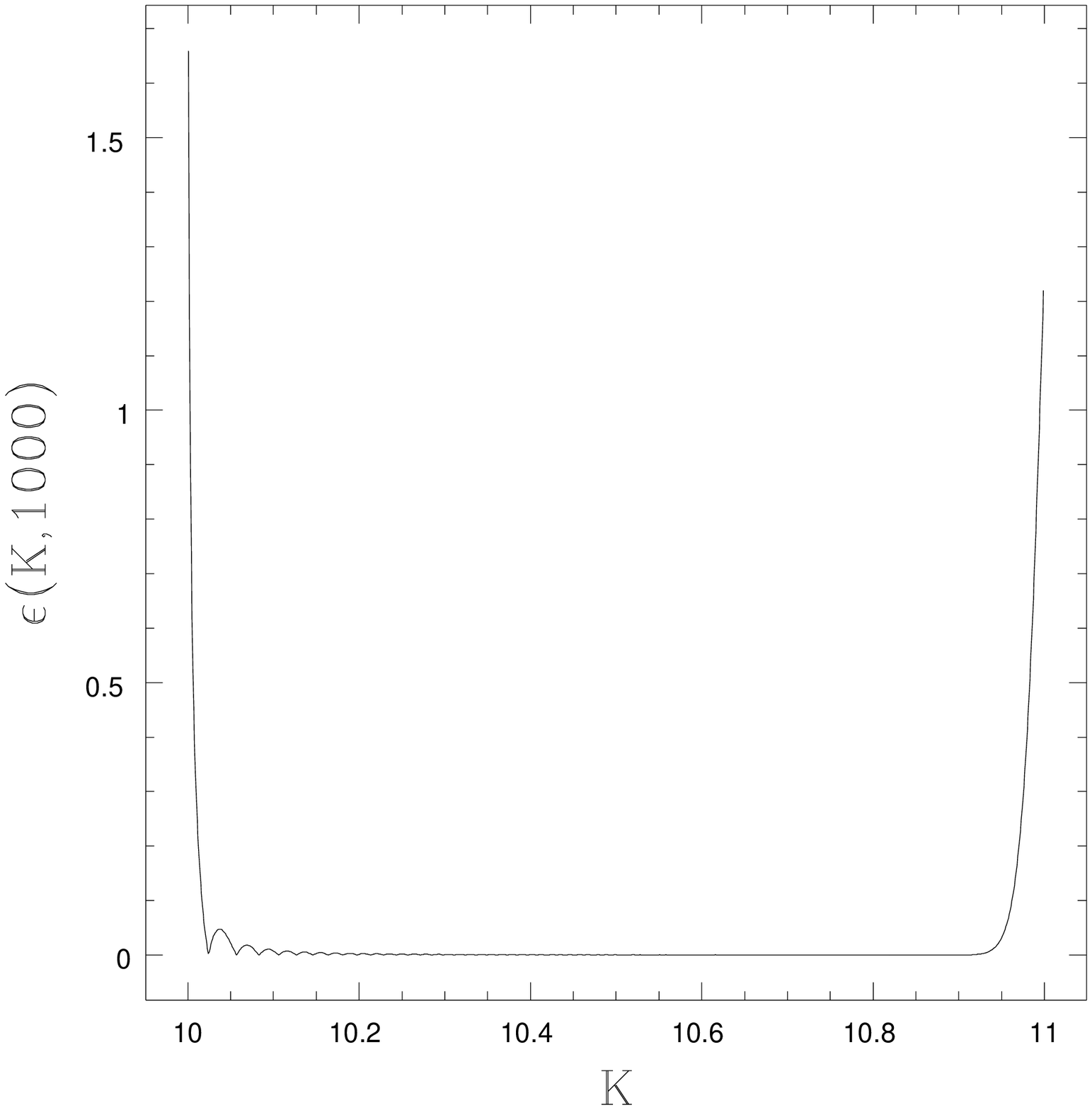}

\vspace*{-8cm}
\hspace*{9cm}
\epsfxsize 8cm
\epsfysize 8cm
\epsfbox[18 150 590 716]{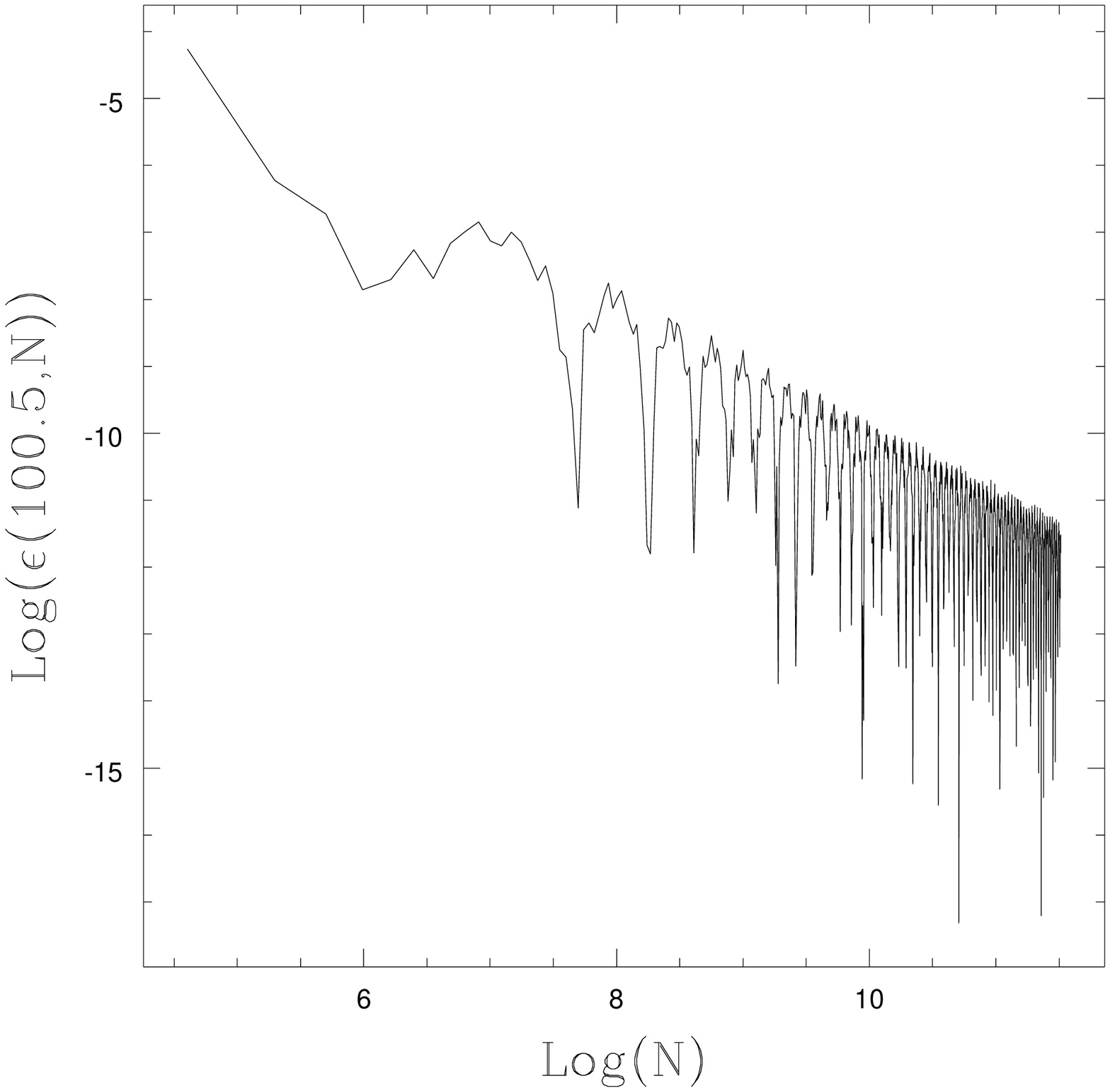}

\caption { The error for the standard map on the torus 
as a function of (a, left) $k$, with $N=1000$;
(b, right) inverse Planck's constant
$N$, with  $k=100.5$}
\end{figure}
\vskip 3mm

In Fig. 4(b) we show the absolute value of the error $\eps\,=\,
|\mbox{Tr}(U)-\mbox{Tr}(U)_{sc}|$ for a fixed value of $k=100.5$, and
a range of $N$. It is clear that the errors involved are not
monotonically decreasing with $N$, but that there are fluctuations in
the mean. The mean however very nearly follows $N^{-1}$
behaviour. This should be contrasted with the more cleaner but slower
$N^{-1/2}$ behaviour of the sawtooth map.  In general the errors
incurred for the smooth standard map is smaller than those of the
sawtooth map. We see that this is a property that is carried over from
the maps on the cylinder. It is unclear whether the standard map fairs
better due to its smoothness or due to some other property. This said,
we note that the trace formula approximation behaves better with 
increasing chaos in the case of the sawtooth map than the standard map.

At a fixed value of $N$ the behaviour of the trace for increasing $k$
is quite complex. No simple rule like, higher chaos implies better
semiclassical accuracy emerges. First we note that the ratios or the
relative error fair quite poorly, with large abrupt deviations.
Fig. 5(a) shows the relative errors
$|\eps|/|\mbox{Tr}(U)_{sc}|$, but the range in the $y$ axis has been
lessened to highlight the predominant features. The error could become
as high as 40 in the range of $k$ shown.  This is however easy to
understand in the context of the Eq. (7) for the map on the cylinder,
as due to the mismatch of the zeros of the semiclassical and the
quantum traces. We expect the absolute value of the errors to fair
better. Indeed when $N$ is small, one observes an increase of accuracy
with $k$, and this is a power law for small $N$. Thus we show in
Fig. 5(b) the logarithm of the errors with the logarithm of
$k$, and we observe the rule that $\eps =C k^{-\gamma}$, with $\gamma
\sim .25$, for the extreme quantum case $N=2$. Here the $k$ go in
steps of unity with their fractional part being uniformly .5.

\begin{figure}[tbh]

\vskip 2mm
\hspace*{10mm}
\epsfxsize 8cm
\epsfysize 8cm
\epsfbox[18 150 590 716]{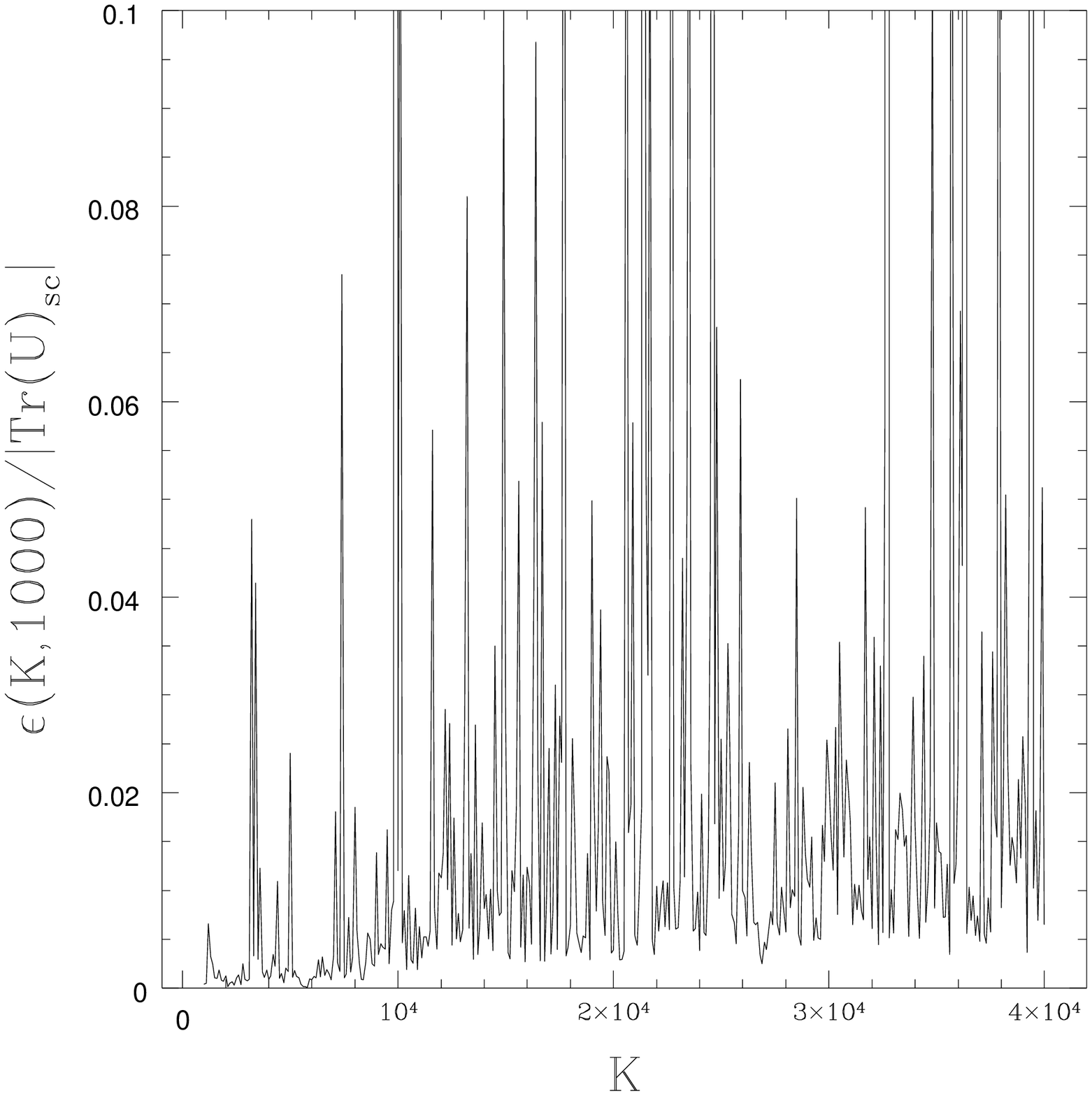}

\vspace*{-8cm}
\hspace*{9cm}
\epsfxsize 8cm
\epsfysize 8cm
\epsfbox[18 150 590 716]{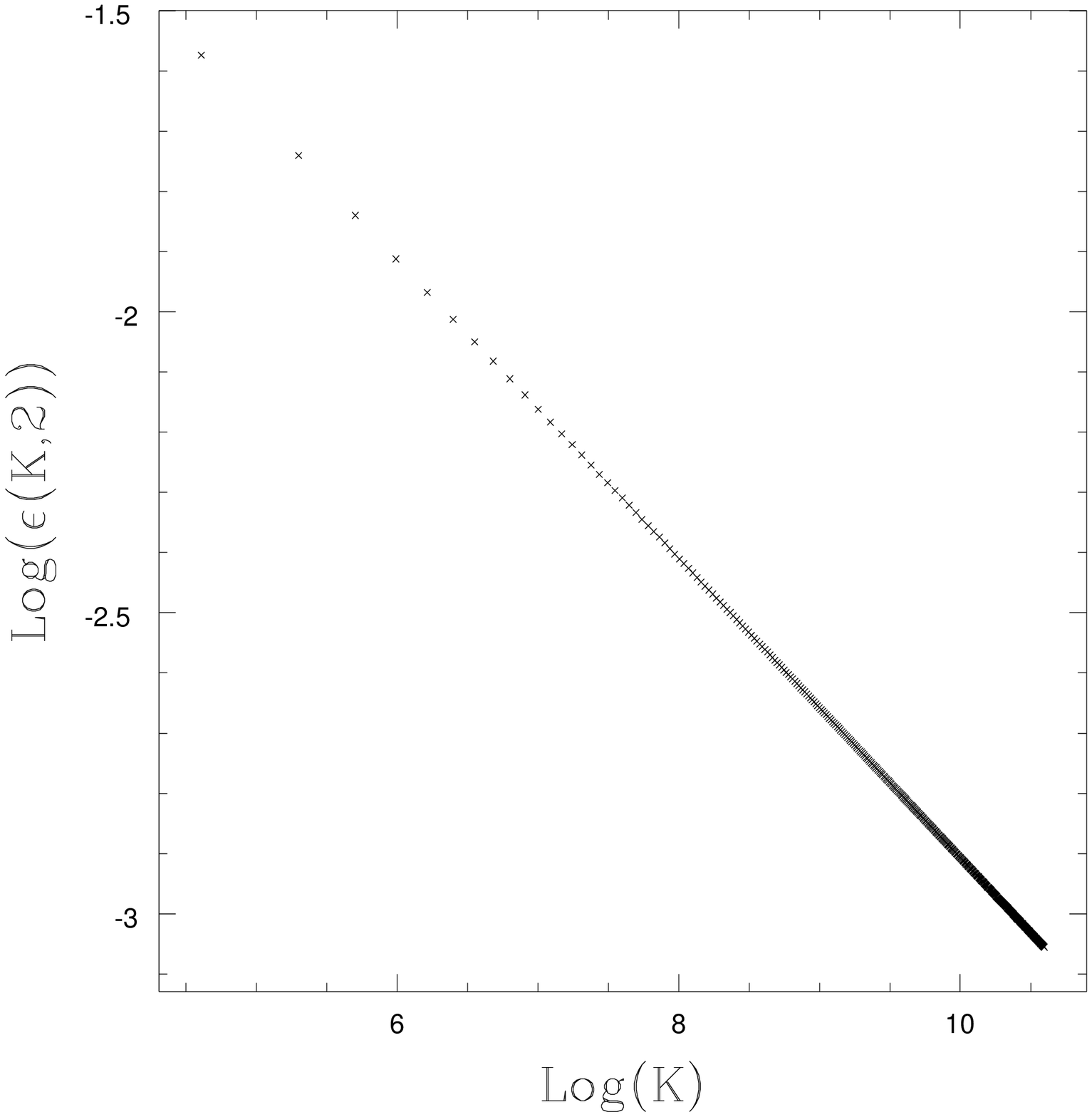}

\caption{(a, left) The relative error for the standard map, 
as a function of the chaos parameter $k$, $N=1000$; 
(b, right) the absolute error as a function of $k$ in the extreme 
quantum limit, $N=2$.}

\end{figure}
\vskip 3mm

However as $N$ is increased we observe that this is no longer true,
and that the errors tend to oscillate with $k$ in a fairly complex
manner. The overall tendency seems to be in fact a deterioration of the
accuracy with $k$. Fig. 8 shows the errors as a function of $k$ in a
large range. The interesting aspect is that the oscillation period
seems to increase geometrically. Therefore there is a possibility that
the errors scale with $k$, but in a complex manner. However increasing
$k$ even higher seems to render the errors an almost monotonic growth,
and we cannot comment on the many details until further analytic
estimates are made and computer generated errors are under control.

\begin{figure}[tbh]

\vskip 2mm
\hspace*{1cm}
\epsfxsize 15cm
\epsfysize 8cm
\epsfbox[18 150 590 716]{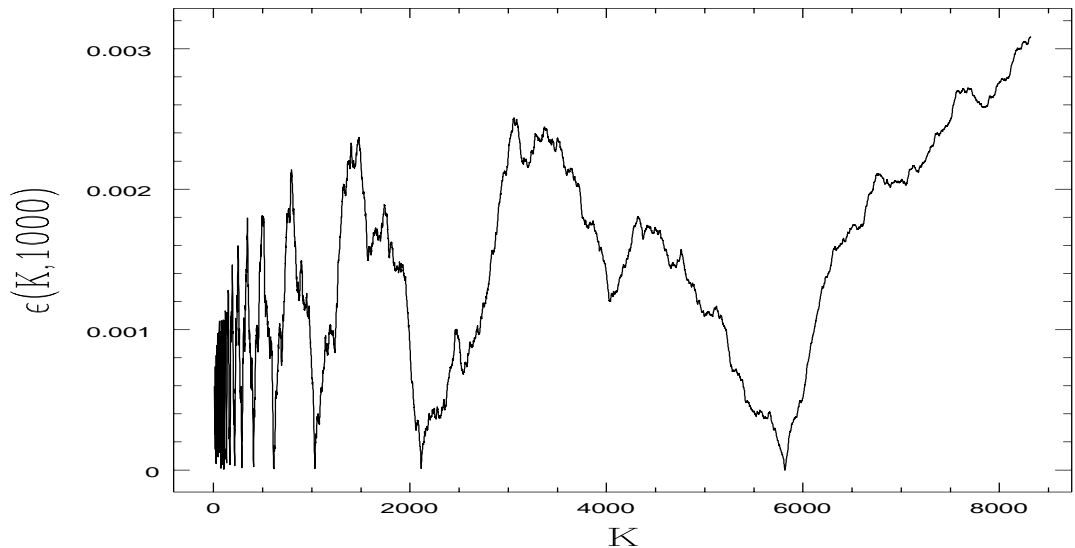}

\caption{The error as function of $k$ in a wide range 
of $k$, for the standard map with $N=1000$.}

\end{figure}
\vskip 3mm

An important difference between the semiclassics of the sawtooth map
and the standard map is that because of the piecewise linear nature of
the sawtooth map there was no need to truncate Taylor expansions about
the periodic orbits, while in the case of the standard map this
truncation is done up to second order. Higher order corrections can be
worked out and have formed the subject of recent works [15,16], it
should be interesting to compare the errors when such higher order
terms are included.

\section{Summary}

Using maps as models of quantum chaos we have studied the simplest
traces from the point of view of semiclassical approximations, hoping
that this will reflect some more general aspects of errors involved in
the Gutzwiller's periodic orbit sum or trace formula. We found that
indeed the maps on the cylinder had controllable relative errors. We
found that the smooth standard map faired better than the sawtooth map
in the order of magnitude of the errors. We also noted that the
mismatch of zeros between the semiclassical and the quantum can lead
to deterioration of relative errors. In the case of toral maps we
displayed a rather remarkable fact that even in the presence of fixed
points on discontinuities the trace formula may be exact at the level
of the time one trace. Otherwise we showed that more generally there
is no $\log(\hbar)$ divergences in the sawtooth map, unlike the case
of the quantum bakers map. We also noted that the error may actually
decrease with increasing chaos, a rather counter intuitive fact, which
has quantitative evidence in the case of the sawtooth map. However we
caution that this cannot be a general principle as a study of the
standard map has shown a much more complex behaviour of the error. It
is indeed a matter of importance to study and contrast the two maps,
and understand why the errors behave qualitatively differently.

We have restricted our study to the time one trace and indeed it is natural
and necessary that we study the higher traces. 
For instance to study the errors involved in the semiclassical 
evaluation of $\mbox {Tr} (U^{2})$  leads us into the problem
of enumerating period two orbits, which can be fairly easily computed in the
cases studied, but which are much larger in number than the fixed points.
Preliminary results in this direction indicate that for the sawtooth map
the error behaviour of the time two trace is similar to that of the trace,
although the errors are larger. We have also displayed some remarkable
curves akin to curlicues that arise out of a study of the trace of the 
standard map on the torus.

\newpage
\vspace{1.3cm}
\begin{center}
 {\bf References}
\end{center}

\begin{itemize}

\item [{[1]}] M.C. Gutzwiller, ``Chaos in Classical and Quantum Mechanics'',
 Springer (New York, 1990).

\item [{[2]}] M.A. Sepulveda, S.Tomsovic and E.J. Heller,  Phys.
Rev. Lett. {\bf 69} (1992), 402.

\item  [{[3]}] N.L. Balazs, and A. Voros, Ann. Phys., (N.Y.){\bf 190}
(1989), 1.

\item [{[4]}] P.W.O' Connor, S. Tomsovic, and E.J. Heller, Physica {\bf 55D}
 (1992), 340;
 A.M.O. De Almeida, and M. Saraceno, Ann. Phys., (N.Y.)
{\bf 210} (1991), 1;
M. Saraceno,  Ann. Phys., (N.Y.) {\bf 199} (1990), 37.

\item [{[5]}] M. Saraceno, and A. Voros, Physica {\bf 79D} (1994), 206.

\item [{[6]}] A. Lakshminarayan, Ann. Phys., (N.Y.) {\bf 239} (1995), 272.

\item [{[7]}]  F. M. Izrailev, Phys. Rep. {\bf 196} (1990), 299.

\item [{[8]}] G. N. Watson, "A Treatise on the Theory of Bessel Functions'',
 Cambridge University Press (London, 1966). 

\item [{[9]}] J.H. Hannay, and M.V. Berry, Physica {\bf 1D} (1980), 267.

\item [{[10]}] A. Lakshminarayan, Phys. Lett. {\bf A192} (1994), 345; 
A. Lakshminarayan and N. L. Balazs, Chaos, Solitons, Fractals (Osaka),
 {\bf 5} (1995), 1169.

\item [{[11]}] J. Ford, G. Mantica, and G.H. Ristow, Physica
{\bf 50D} (1991), 493.

\item [{[12]}] J. P. Keating, Nonlinearity {\bf 4} (1991), 335.

\item [{[13]}] M. V. Berry, and J. Goldberg, Nonlinearity {\bf 1} (1988), 1.

\item [{[14]}] J. M. Greene, J. Math. Phys. {\bf 20} (1979), 1183.

\item [{[15]}] P. Gaspard, and D. Alonso, Phys. Rev. {\bf
A47} (1993), R3468.

\item [{[16]}] G. Junker and H. Leschke, Physica {\bf D56} (1992), 135.

\item [{[17]}] M. Tabor, Physica {\bf D6} (1983), 195.

\end{itemize}

\newpage
\begin{center}
{\bf Appendix}
\end{center}
Consider the following map defined on the unit torus: 
\beq
	\begin{array}{ccl}
	q_{j+1} & = & q_{j} \, +\, p_{j} -l_{j}\\
	p_{j+1} &=& p_{j}\, -\, V^{\prime}(q_{j+1}) - m_{j},
	\end{array}
\eeq
where $V(q)$ is a periodic ``kicking'' potential with unit
periodicity, the prime indicates derivative, and the integers 
$m_{j}$ and $l_{j}$ are such that
the map is restricted to the unit torus (winding numbers). 
They are the ``modulo one'' 
operations. As is well known the above map can be derived from a 
time dependent Hamiltonian system. Also there exists a 
generating function for the map on the torus given by
\beq
S(q_{j}, q_{j+1}; l_{j}, m_{j})\,=\, \frac{1}{2}(q_{j+1}-q_{j}+l_{j})^{2}
-V(q_{j+1}) - m_{j} q_{j+1}
\eeq
from which the map can be derived as $p_{j}\,=\,- \partial
S/\partial q_{j}$ and $p_{j+1}\,=\, \partial S /\partial q_{j+1}$.

The quantum map corresponding to the Eqs.(26) is given by the finite
dimensional unitary matrix
\beq
\br n |U|n^{\prime} \kt \,=\, \frac{e^{-i \pi/4}}{\sqrt{N}} \exp
(\frac{i \pi}{N} (n-n^{\prime})^{2} \,-\, 2 \pi i N V(\frac
{n}{N})).	
\eeq
This takes on the function of the propagator as it connects
states after consecutive kicks.
Here the representation is in the position basis and $n$
and $n^{\prime}$ may take the integer values from 0 to $N-1$.
$N$ is itself the dimensionality of the Hilbert space and is
also the inverse Planck constant, and is restricted to be even
integers for preservation of toral boundary conditions. Semiclassics means the
study of the unitary operator as $N \rightarrow \infty$.
 
The object of primary interest below is the trace of the powers
of the propagator from which the spectrum may be derived by a
Fourier transform. This can be written as
\beq
\mbox{Tr}(U^{T})\,=\,\frac{e^{-i \pi T/4}}{N^{T/2}}
 \sum_{  
n_{i} = 0 }^{N-1} \, \exp\left( \frac{2 \pi i}{N} 
\sum_{j=1}^{T} \left( n_{j}^{2} \, -\  n_{j} n_{j+1} \,-\, N^{2} \,
V(\frac{n_{j}}{N})\right) \right), 
\eeq
with \[ n_{T+1}\,=\, n_{1} .\]
It is implied in the above that the outer sum is over the  $T$
variables $n_{i}, \, i=1,2,\ldots,T$. Using the Poisson
summation formula we can rewrite the above as 
\beqa
\mbox{Tr}(U^{T})\,=\,e^{-i \pi T/4}\, N^{T/2}
 \sum_{k_{i} = -\infty }^{\infty} \, \int_{ -\epsilon}^{1
-\epsilon}
\, dx_{1} \dots dx_{T} \nonumber \\ 
\exp \left(2 \pi i  N 
\sum_{j=1}^{T} \left(k_{j} x_{j} \,+\,  x_{j}^{2} \,-\,  x_{j} x_{j
+1} \,-\, V(x_{j})\right)
\right),
\eeqa
with \[ x_{T+1}=x_{1}, \;\mbox{and}\;\;\; k_{T+1}=k_{1}. \]
The stationary phase approximation, assuming that
$N$ is large, gives the following $T$ conditions
\beq
x_{j+1}\,=\, 2 x_{j} \,-\, x_{j-1}\,-\, V^{\prime}(x_{j})\, +\, 
k_{j}; \;\; \; \; j=1,\ldots,T,
\;\;x_{0}=x_{T},\;\;x_{1}=x_{T+1},
\eeq
with the further condition that \[ 0 \, \le \, x_{j} \, < \,
1. \] 
These are precisely the equations that determine the
period $T$  orbits of the  map. The integers $k_{j}$ are related
to the integers $l_{j}$ and $m_{j}$ of Eqs. (26) by the relation
$k_{j}=l_{j-1}-l_{j}-m_{j-1}$. With the  assumption  that they uniquely
determine a periodic orbit of period an integer fraction of $T$,
we restrict the infinite sum to only those that relate to such a
periodic orbit labeled below by $\beta$. This is the first  
approximation and the second
one is that we will extend the ranges of integration to the
whole real line. These approximations together are exact in the case
of the cat maps [9,11,12], and the third related approximation
is that we Taylor expand the potential about the periodic
orbits retaining up to the quadratic terms. This approximation is 
unnecessary for the piecewise linear sawtooth maps [10], but 
is necessary in general. 
 
Therefore we define new variables $y_{j}$ such that
$x_{j}=q_{j}+y_{j}$ and $q_{j}$ is the central periodic orbit.
\beqa
\mbox{Tr}(U^{T}) \, \sim \,e^{-i \pi T/4} N^{T/2} \, \sum_{\beta} 
\, \exp\left(2 \pi i N \sum_{j=1}^{T}
\left(-q_{j}q_{j+1}\, +\,q_{j}^{2}\,+\, q_{j}\,k_{j}-V(q_{j})
 \right)
\right) \; \times \nonumber \\ 
\int_{-\infty}^{\infty} dy_{1} \dots dy_{T} \exp\left(2 \pi i  N \sum_{j=1}^{T}
\left(y_{j}^{2}\,-\,  y_{j}y_{j-1} - y_{j}^{2} \,V^{\prime \prime}(q_{j})/2 
\right)\right)
\eeqa
The sum 
\beq
\sum_{j=1}^{T}
\left(-q_{j}q_{j+1}\, +\,q_{j}^{2}\,+\, q_{j}\,k_{j}-V(q_{j})
\right)
\eeq
can be identified with the action of the periodic orbit, as from
Eq.(27) one gets
\beq
S_{\beta}\,=\, \sum_{j=1}^{T} \,
(q_{j}^{2}-q_{j}q_{j+1}-V(q_{j})+q_{j} k_{j}+l_{j}^{2}/2),
\eeq
which is essentially the  sum in Eq. (33). We can neglect the
term $l_{j}^{2}/2$, from the Lagrangian; this is legitimate as the
action occurs as $e^{2 \pi i N S}$, and since $N$ is an even integer
these terms do not matter. Note that the trace formula at $T=1$ is
easily obtained from the above and has been used in the main body of
this paper.

The integral is  evaluated by  standard methods. 
\beq 
\int_{-\infty}^{\infty} d^{T}y \, \exp(i \pi N y^{t}A_{\beta}y)
\,=\, \frac{e^{i \pi
T/4}}{N^{T/2}}\;\frac{e^{-i \pi \nu_{\beta} /2}}{\sqrt{|\mbox{Det}
(A_{\beta})|}} .
\eeq
Here $y$ denotes the $T$ component vector $\{y_{j}\}$ and
$y^{t}$ is its transpose. The real symmetric matrix $A_{\beta}$ is given by 
\beq
A_{\beta}\,=\, \left( \begin{array}{cccccc}
	2-V^{\prime \prime}(q_{1}) & -1 & 0 & .\; . & .\;.& -1\\
	-1 & 2-V^{\prime \prime}(q_{2}) & -1& 0& .\;.&0\\
	:  & : &:&:&:&:\\
	:&:&:&:&:&:\\
	:&:&:&:&:&:\\
	-1 & 0 & .\;.&.\;.& -1& 2-V^{\prime \prime}(q_{T})
	\end{array}
	\right).
\eeq
$\nu_{\beta}$ is a ``Maslov like'' index, and is the number of
negative eigenvalues of the matrix $A_{\beta}$.

The nontrivial problem of evaluating the determinant is already
solved and the results are well known [14]. The determinant can be
related to the stability of the periodic orbit. If the residue
of the periodic orbit, which is simply related to the trace of
the stability matrix, is $R_{\beta}$, then $R_{\beta}\,=\, -\frac{1}
{4} \mbox{Det}(A_{\beta})$.
or if we denote the stability matrix eigenvalues by
$\lambda_{+}^{\beta}$ and $\lambda_{-}^{\beta}$, Det$(A_{\beta})\,=\, 
\lambda_{+}^{\beta}+\lambda_{-}^{\beta} -2$.

Thus finally we can write the asymptotic periodic orbit sum, or
the trace formula, for the toral maps as
\beq
\mbox{Tr}( U^{T}) \,\sim \, \sum_{\beta}\; \frac{\exp\left(2 \pi i  N S_{\beta}
-i \pi \nu_{\beta}/2 \right)}{\sqrt{|\lambda_{+}^{\beta \, T} +
\lambda_{-}^{\beta \, T} -2|}} .
\eeq
The sum is over periodic orbits whose periods are integer fractions of $T$
are labeled by $\beta$. 
This  is of the canonical Gutzwiller-Tabor form having in the
exponent the {\em action} of the period $T$ orbits, and the prefactor
explicitly depending on the {\em stability } of these orbits. 
The piecewise linear maps such as the baker's map, or the
sawtooth map are such that the Maslov like phases are zero and
the only type of periodic orbits are those of the direct
hyperbolic kind. Now however we can study the vast class of {\it
mixed} systems, such as the well known standard map. It is then
of interest to ask questions such as how long the semiclassical 
approximations are valid in such systems which one knows are
generic. The quantum mechanics on the torus is exact and 
thus we expect that the derivation above will facilitate the
investigation of the periodic orbit sum for mixed systems. 
In this note we have  used the above formula only in the special 
case of $T=1$.

\end{document}